\documentstyle[aps,12pt]{revtex}

\begin{document}

\title{Holographic bound from second law of thermodynamics}

\author{ Jacob D. Bekenstein\thanks{E--mail:
     bekenste@vms.huji.ac.il}}

\address{\it The Racah Institute of Physics, Givat Ram, Jerusalem 91904,
Israel}

\date{\today}
\tighten
\maketitle

\begin{abstract}

A necessary condition for the validity of the holographic principle is
the holographic  bound: the entropy of a system is bounded from above by a
quarter of the area of a circumscribing surface measured in Planck areas. 
This bound cannot be derived at present from consensus fundamental theory. 
We show with suitable {\it gedanken\/} experiments that the holographic bound
follows from the generalized second law of thermodynamics for both generic
weakly gravitating isolated systems and for isolated, quiescent and
nonrotating strongly gravitating configurations well above Planck
mass.  These results justify Susskind's early claim that the holographic bound
can be gotten from the second law.
\end{abstract}
\subsection*{1. Introduction}
The influential holographic principle\cite{thooft,susskind} relates the
physical content of a theory defined in a spacetime to that of another theory
defined only on the spacetime's boundary.  This principle is viewed as a
guideline to the ultimate physical theory. A consistency requirement on it is
that the boundary of any system should be able to encode as much information
as required to enumerate and describe the quantum states of the bulk system. 
In view of the correspondence between information and entropy, and the example
of black hole entropy, this requirement has been translated into the
holographic bound (HB): ``the  entropy of a system is bounded from above by a
quarter of the area of a circumscribing surface measured in units of
the squared Planck length, $\ell_P{}^2\equiv G\hbar/c^3$'' (in the sequel we
set $c=G=1$ and eliminate $\hbar$ in favor of $\ell_P{}^2$; Planck's mass is
also denoted by $\ell_P$).

Independently of the above, the existence of {\it some\/} bound on the entropy
of a closed system is always taken for granted in thermodynamics.  Since the
second law requires the entropy of a closed physical system to increase in
most transformations, it is usually inferred that that entropy must 
eventually reach a maximum. Obviously this inference takes it for granted
that the entropy of any closed system is bounded from above.  Such bound is
not always in evidence; for example, in Newtonian gravity theory there seems
to be no upper bound to the entropy of a gravitationally bound system of
given mass and energy\cite{tremaine}.  Of course, we know that gravitation is
better described by general relativity; this is, in fact, what makes a bound
on the entropy of a gravitating system feasible. When one turns
from a coarse--grained description of matter, as assumed above, to a
microscopic one, the entropy of a closed system does not change (Hamiltonian
classical evolution or unitary quantum evolution).  However, one can still
ask whether the quantum entropy here is subject to some bound for given
exterior constraints.  A generic bound on entropy would have weighty
implications; for example, it would restrict the maximum information that
could be stored in a system whatever the future technology.  

One of the early consequences of the ascription of entropy to black holes,
and the formulation of the generalized second law (GSL) for systems
containing black holes\cite{bek72}, was the realization\cite{bek81,bek94}
that a universal entropy bound valid for all systems - black hole or not -
can be stated solely in terms of the circumscribing radius and proper energy of
the system [see Eq.~(\ref{uni_bound}) below].  The argument was that the
second law would be violated if a system violating the universal bound
were accreted by a black hole.  With the emergence of the HB the closely
related question comes up:  can one also infer the HB from the GSL ? 
Susskind\cite{susskind} has early argued that this is so from a {\it
gedanken\/} experiment in which a system violating the HB is forced to
collapse into a black hole by the addition of extra entropy--free matter.
Susskind interprets the ensuing apparent violation of the GSL as evidence
that the envisaged system cannot really exist.

However, Susskind's argument contains a loophole, even in a clearer
reformulation\cite{wald}.  One imagines the system as a spherically symmetric
one of radius $R$, energy $E$ (with $R>2E$, of course) and  entropy ${\cal
S}$ with ${\cal S}>\pi R^2/\ell_P{}^2$.  A spherically
symmetric and concentric shell of mass $R/2-E$ is dropped on the system; by
Birkhoff's theorem the total mass is now $R/2$.  {\it If\/} the outermost
surface of the shell reaches Schwarzschild radial coordinate $r=R$, the
system becomes a black hole of radius $R$ and entropy $S_{BH}=\pi 
R^2/\ell_P{}^2$, which is lower than the original entropy ${\cal S}$ ! 
Wald concludes from this, not that ${\cal S}>\pi
R^2/\ell_P{}^2$ is forbidden (HB is required), but rather that if ${\cal
S}>\pi R^2/\ell_P{}^2$, the outcome of the process is not a black hole,
e.g. the shell could bounce.  This is in line with his position that the GSL
works without regard to entropy bounds on matter\cite{UW}. However,  as we
show below, alternative {\it gedanken\/} experiments (related to the ones
introduced in Ref.\cite{MG7} and the third reference in\cite{bek72}) allow one
to infer, without ambiguity, the HB from the GSL for two broad
classes of systems.  Thus we generally justify Susskind's view.

Before starting let us mention that there do exist violations of the
HB as here stated.  For example, a spherically symmetric mass already inside
its own Schwarzschild radius eventually violates it because the enclosing
area can get as small as required, while the enclosed entropy can only
grow\cite{bousso}.  So does a sufficiently large spherical chunk of a flat
Robertson--Walker universe because its enclosing area grows like radius
squared while the enclosed entropy does so like radius cubed.  Both these
examples feature strongly self--gravitating and dynamical systems.  The
second refers to an unisolated system.  Bousso\cite{bousso} has provided a
broader HB--like bound free from such failures (a rather different entropy
bound has lately been proposed by Brustein and Veneziano\cite{BV}).  Bousso's
bound can, however, be intricate to apply in cases which lack high
symmetry\cite{ellis}.   Thus the original HB, which is simpler to apply, would
still be very valuable if its range of validity could be delineated in a
systematic way.  Here we use the GSL to show that it can be trusted
for generic weakly self--gravitating isolated systems as well as for
quiescent, nonrotating, isolated strongly self--gravitating systems which are
massive on Planck's scale. 
\subsection*{2. Weakly gravitating systems}

Consider a weakly self--gravitating system ${\cal U}$ of arbitrary structure
and constitution having proper mass--energy $E$ and bearing entropy ${\cal
S}$.  For convenience ${\cal U}$ can be enclosed in an entropy--less snug
spherical box of radius $R$ concentric with ${\cal U}$'s center of mass; we
include its mass in $E$. This sphere will be dropped freely from rest and from
far away into a Schwarzschild black hole of mass $M$ satisfying $M\gg E$ and
$2M\gg R$. ${\cal U}$'s global energy is thus very near $E$.  Obviously
${\cal U}$ will not be torn up by tidal forces as it approaches the black
hole.  Also, no reason is known why the black hole should be destroyed when
the much smaller and lighter system plunges into it.  Hence we conclude that
the black hole absorbs ${\cal U}$ and increases its mass by $E$ in the
process.  We are here ignoring energy losses to gravitational radiation, known
to be of $O(E^2/M)$\cite{davis}, as well as losses to Hawking radiation during
${\cal U}$'s infall; we shall justify this last omission presently.  

The black hole's entropy, originally $S_{BH} = 4\pi M^2/\ell_P{}^2$, thus
increases by $\Delta S_{BH}=8\pi ME/\ell_P{}^2 +O(E^2)$.  By the GSL $\Delta
S_{BH}$ plus the radiation entropies are at least as large as
${\cal S}$.  We neglect the radiation entropies of both sorts because the
corresponding energies are negligible.  We thus have
\begin{equation}
{\cal S}\leq {8\pi ME\over \ell_P{}^2}.
\label{raw_bound}
\end{equation}
Thus far we have only required $M\gg E$ and
$2M\gg R$.  Let us settle on the choice  $M={\scriptstyle 1\over 
\scriptstyle 8}R^2/E$.  This satisfies both our requirements provided
$2E\ll R$, i.e., provided ${\cal U}$'s self--gravity is weak.   With our
choice for
$M$  Eq.~(\ref{raw_bound}) gives
\begin{equation}
{\cal S}\leq {\pi R^2\over \ell_P{}^2},
\label{holographic}
\end{equation}
which is precisely the HB appropriate to ${\cal U}$.  We stress
it applies to a system of generic structure and composition provided its
self--gravity is weak. In the sequel we extend this result to a large class
of strongly self--gravitating systems.  

Let us now justify the neglect of energy losses to Hawking's radiance.  One
can approximate the power emitted by a Schwarzschild black hole by that of a
sphere with radius $2M$  radiating according to the Stefan--Boltzmann law
at the Hawking temperature $T_H= \ell_P{}^2/(8\pi M)$.  Thus 
\begin{equation}
\dot M \equiv {dM\over dt}\approx - {N\ell_P{}^2\over 15360 \pi M^2},
\label{power}
\end{equation} where $N$ is
the effective number of radiated species (photons contribute unity to
$N$). If  ${\cal U}$ was dropped at Schwarzschild time $t=0$ when its center of
mass was a Schwarzschild radial coordinate $r=\alpha 2M\ (\alpha\gg 1)$,
then the  bottom of the sphere reaches the horizon $r=2M$ (the sphere's
center is at {\it proper\/} height $R$ above it), and the infall is finished
for all practical purposes, at  time
\begin{equation}
t \approx 4M\left[{\scriptstyle 1\over \scriptstyle 3}\alpha^{3/2}
-{\scriptstyle 1\over \scriptstyle 4}\ln(R/4M)+O(\alpha^{1/2})\right].
\label{time}
\end{equation}
Eq.~(\ref{time})  is obtained by integrating exactly the radial
component of the geodesic equation for ${\cal U}$'s center of mass
in the Schwarzschild metric, and dropping subdominant terms.

The energy ${\cal E}$ emitted during the infall to the black hole of mass
$M={\scriptstyle 1\over  \scriptstyle 8}R^2/E$ is thus
\begin{equation}
{\cal E}=|\dot M|\times t \approx {N\over 480\pi}{\ell_P{}^2\over
R^2}\left[{\scriptstyle 1\over
\scriptstyle 3}\alpha^{3/2}-{\scriptstyle 1\over \scriptstyle 4}\ln(2E/R)
\right] E,
\label{energy}
\end{equation}
The very small factor $\ell_P{}^2/(480\pi R^2)$, $10^{-40}$ for a proton and
much smaller for macroscopic systems, easily outbalances the assumed large
$\alpha^{3/2}$ (say $10^9$), any reasonable finite $N$ (say $10^{4}$), and
the logarithmic factor (say for $2E/R\sim 10^{-10}$).  Thus ${\cal E}\ll E$
and hence it was justified to equate the change in black hole mass with $E$. 
Similarly, since the entropy of the Hawking radiation is $\sim {\cal E}/T_H$,
it can safely be neglected in comparison with
$\Delta S_{BH} =E/T_H$.  The fact that ${\cal E}\ll E\ll M$
justifies {\it a posteriori\/} our use of the Schwarzschild metric for the
black hole's exterior.

Another effect of Hawking's radiance that we have neglected so far is
radiation pressure on the infalling sphere.  One might fear that the
consequent change in the postulated geodesic motion might drastically
prolong the infall time, or perhaps even prevent  ${\cal U}$ from reaching the
horizon.   We now show that neither of these possibilities materializes. 
Obviously the momentum flux at radial coordinate $r$ is $|\dot M|/(4\pi
r^2)$.  The rate at which momentum is received from the radiation by the
infalling sphere, in its own rest frame (four velocity $u^\beta$), is $|\dot
M|/(4\pi r^2)$ times the sphere's crossection $\pi R^2$, times $(dt/d\tau)^2$,
where $\tau$ is the sphere's proper time.  One factor $dt/d\tau$ accounts
for the blueshift of momenta of the Hawking quanta perceived in the
falling sphere's frame;  the second corrects for the faster arrival of
quanta due to the time dilation and gravitational redshift.  If the sphere
reflects radiation well, one has to multiply this result by a factor
between 1 and 2 to account for backscattering, but this is a trivial
correction here.

Since the sphere falls from $r\gg 2M$, we have $dt/d\tau\approx(1-2M/r)^{-1}$. 
Putting all the factors together with Eq.~(\ref{power}) and dividing by
$E$,  we find the acceleration of the falling sphere measured in its
own frame:
\begin{equation}
a\approx{N \ell_P{}^2 R^2\over 61440  M^2 Er^2}{1\over (1-2M/r)^2}
\label{acceleration}
\end{equation}
This is to be compared with $g = Mr^{-2}(1-2M/r)^{-1/2}$, the
acceleration scalar of a stationary point at radius $r$.  Then
Eq.~(\ref{acceleration}) gives
\begin{equation}
{a\over g}={N \ell_P{}^2/E\over 7680  R}\left[{R\over
2M(1-2M/r)^{1/2}}\right]^3.
\label{acceleration_ratio}
\end{equation}
Since the Compton length $\ell_P{}^2/E$ of a system is always smaller (and can
be much smaller) than its radius, and $R\ll 2M$ in our {\it gedanken\/}
experiment, we find that the acceleration due to radiation pressure is
negligible compared to the natural scale $g$ as long as the sphere is not
very close to the black hole's horizon.  It is doubtful whether as
$r\rightarrow 2M$, the Hawking radiance continues to be felt in the freely
falling frame.  At any rate, because $R\ll 2M$, the factor in square brackets
in Eq.~(\ref{acceleration_ratio}) only grows to 2 when the bottom of the
sphere touches the horizon.  Hence the radiation pressure deceleration is
totally negligible.  We conclude that ${\cal U}$'s center of mass does move
accurately on a timelike geodesic throughout its fall, as has been assumed
all along.  Smallness of $a/g$ also means that it is unnecessary to correct
for quantum buoyancy effects, as is the case when the system is
suspended\cite{UW,bek94}.

The effects which might have spoiled the derivation of the HB are so utterly
negligible that we can improve on that bound by taking $M$ smaller than
${\scriptstyle 1\over \scriptstyle 8}R^2/E$ in the raw
bound (\ref{raw_bound}).  We obviously have to stop this optimization when $2M$
is at least a few times $R$, so that the system can fit whole in the black
hole.  We thus get
\begin{equation}
{\cal S}\leq {4\zeta\pi ER\over \ell_P{}^2},
\label{uni_bound}
\end{equation}
where $\zeta=O(1)$.  Eq.~(\ref{uni_bound}) is the universal entropy
bound\cite{bek81}.  More careful analysis by a different
route\cite{bek81,bek94} shows that one can take $\zeta={\scriptstyle 1\over
\scriptstyle 2}$.  In this form the bound applies also to equilibrium black
holes: for every Kerr--Newman black hole, the ``radius'' $R$ of the horizon
(with $4\pi R^2$ equal to the horizon area) is no bigger than
twice the black hole's mass $E$, while $S_{BH}=\pi
R^2/\ell_P{}^2$.  Further, one may obviously derive the HB directly from bound
(\ref{uni_bound}) by simply inserting the weak self--gravity condition $2E\ll
R$\cite{MG7}.  Thus for weakly self--gravitating systems the universal entropy
bound is stricter than the HB\cite{MG7,BV}.

\subsection*{3. Strongly gravitating systems}

When the system's self--gravity is no longer weak ($E \sim R$), the condition
$M={\scriptstyle 1\over \scriptstyle 8}R^2/E$ implies that $E$ is no longer
small compared to $M$ (system not a small perturbation on
the black hole) and that $R\sim 2M$ (system cannot fall whole into
the black hole).  We thus prefer to derive the HB from the GSL by the following
alternative argument.  

We assume that the system  ${\cal V}$ containing entropy ${\cal S}$ is
isolated, quiescent (does not significantly change in ``time'', e.g. is not
collapsing), and nonrotating, and that the spacetime geometry is
asymptotically flat.   We shall not assume spherical symmetry; however, the
strong self--gravity in the absence of rotation should make the system
nearly spherical.  The quiescence assumption implies that there is an
approximate time symmetry in the problem which allows us to define a time
coordinate, $t$, with the familiar properties and coinciding at infinity with
the usual time (this $t$ may no be unique except in the exactly stationary
case). A tube with topology $S^2\times R$ enclosing the system's world history
can be sliced at specific values of $t$. The area ${\cal A}$ of a typical
slice must  substantially exceed $4\pi(2E)^2$, where $E$ denotes
${\cal V}$'s energy as measured at infinity. For if ${\cal A} \approx
4\pi(2E)^2$, the system would be a black hole.  On the other hand, there is
much evidence that a system whose radius exceeds twice its mass, but not by
much, is gravitationally unstable, i.e., not quiescent as here assumed.  Thus
${\cal A}$ is substantially larger than $4\pi(2E)^2$.  On the basis of these
remarks and our assumptions, we shall now show from the GSL that ${\cal
S}<{\cal A}/(4\ell_P{}^2)$.

Let us enclose ${\cal V}$ in a robust quasispherical box of radius
$r_0\approx 10^2\times  (2E)$ (eventually we shall be able to regard it as
exactly spherical).  The box's interior is at some redshift with respect to
infinity; we measure all energies here in the interior frame.  Obviously
because ${\cal V}$ is strongly self--gravitating, it is small compared to
$r_0$ in spatial extent. From near the box's wall let us drop radially
towards the center of mass of ${\cal V}$ a Schwarzschild black hole of mass
$m$ with  ${\rm Max}[10\ell_P{}^2/\mu, 10(N\ell_P{}^2 E^3)^{1/5}]< m<10^{-1}
E$, where $\mu$ is the typical mass of the lightest massive elementary
constituent of ${\cal V}$, e.g. electron mass for ordinary matter.  The
inequalities are consistent  provided $\mu E> 10^2\ell_P{}^2$ and $E>10^5
N^{1/2}\ell_P$. Because the lightest constituents must be light on the Planck
scale, our argument obviously applies only to systems which are fairly
massive on Planck scale.   The Hawking evaporation timescale ${\cal T}$ of
the black hole is gotten by integrating Eq.~(\ref{power}): ${\cal T} =
5120\pi m^3/(N\ell_P{}^2)$.   In light of the restriction $m>10 (N\ell_P{}^2
E^3)^{1/5}$,
\begin{equation}
{\cal T}>10^{9}(E/m)^2 E.  
\end{equation}
According to Eq.~(\ref{time}), the hole takes time $\sim 10^3E$
to fall to ${\cal V}$.  It may immediately interact strongly, give up
energy and be gravitationally trapped, or it may pass through and out of
${\cal V}$ to rise to near the box wall, fall back, and again pass through
${\cal V}$. During $(E/m)^2$ such passes (and time $\sim 10^3 (E/m)^2
E\ll {\cal T}$) the hole's crossection $4\pi m^2$ would have swept a
volume $\sim 4\surd\pi E^2{\cal A}^{1/2}$ through ${\cal V}$ which is
comparable to that of the whole of ${\cal V}$.  We thus expect the black hole
to be captured by ${\cal V}$ earlier, and with insignificant loss of $m$ to
Hawking radiance. Immediately after this the combined system (also denoted
${\cal V}$) has mass a bit smaller than $E+m$, principally because of
gravitational waves (from the oscillations) which escape from the
box;calculations\cite{davis} make it clear that the radiated energy is only
some fraction of $m$.

Since ${\cal V}$ is strongly self--gravitating, the trapped hole will at
first move at speed close to that of light traversing ${\cal V}$ in
time $\sim E \ll {\cal T}$, but may eventually come to rest at the bottom of
the system's gravitational potential well.  At any rate, the hole will swallow
structures that approach it which are smaller than itself, while its tidal
field will break up the somewhat larger structures into smaller parts, some
down to the scale of elementary constituents.  Since the Compton lengths of
every one of these is smaller than the hole ($\ell_P{}^2/\mu\ll m/10$), it can
swallow anything that comes within its reach.  Hawking radiation pressure
cannot stop the accretion; Eq.~(\ref{acceleration_ratio}) shows that for any
structure whose dimension $R$ exceeds its Compton length, the hole's
gravitation dominates the pressure force unless $N$ is very large. Since a
particular elementary constituent interacts only with certain sorts of
radiation, the effective $N$ that must be used in
Eq.~(\ref{acceleration_ratio}) is never large.  The black hole thus accretes,
grows and becomes more effective at accreting while at the same time cooling
and emitting Hawking radiation slower.  Of course, the accretion will generate
heat, and radiation (e.g. electromagnetic but little gravitational because of
its poor coupling to matter) will leak out of ${\cal V}$ into the box,
supplementing, and perhaps even dominating, the Hawking radiation trapped in
it.  

The black hole's mass cannot decrease by Hawking radiance.
According to Boltzmann's formula [${\rm energy\ density}=N\pi^2
T^4/(15\hbar^3)$], the energy of $N$ species of thermal radiation  filling
the box's volume $ 4\pi (2\cdot 10^2 E)^3/3$ (most of the box is nearly
Minkowskian) at the Hawking temperature $T=\ell_P{}^2/(8\pi m)$ would
be smaller than $10^{-3}m$ for any $N$ if indeed $m>10(N\ell_P{}^2
E^3)^{1/5}$.  Thus backreaction  from the box will prevent the black hole
from losing more than a tiny fraction of $m$ before accretion has made it gain
mass.  This also means the black hole will radiatively equilibrate with the
box in a time much shorter than ${\cal T}$. There is one exception to this
equilibration.  Thermal Hawking gravitons emitted by the hole interact too
weakly with any matter in the box to be confined by it\cite{smolin}; they
stream out freely.  Were the hole not to accrete, it would slowly and
inexorably evaporate due to thermal graviton losses; the graviton emission
timescale  would, of course, be about $N{\cal T}$ (out of $N$ species of
quanta, one is a graviton).  But because of the cooling of the black
hole, the true timescale gets progressively longer.

Given enough time ${\cal V}$ will be fully digested by the hole.  
In the early stages of the process sketched, when the hole is
still small, the evolution is slow and ${\cal V}$ remains quiescent, so
that it emits little gravitational radiation.   The duration $t_s$ of
this stage should be a function of the important scales
$E$ and $m$.  Were $m$ comparable with $E$, the slow stage would obviously be
over in time $\sim E$.  Thus on dimensional grounds we guess $t_s\sim
E(E/m)^\beta$.  For small $m/E$ the accretion effectiveness should scale
as the hole's crossection; hence we deduce that $\beta=2$, so for all
$N$, $t_s\ll{\cal T}$.  Towards the end of the digestion ${\cal V}$ is likely
to evolve rapidly.  Fragments that get ejected from ${\cal V}$  as it is being
digested remain trapped in the box,  and will eventually fall back onto the
hole, be broken up and swallowed.  This last stage should last a time $\sim E$
because it obviously involves an instability, instabilities grow
exponentially, and our two scales $E$ and $m$ are merging.  Overall, the
process of conversion of ${\cal V}$ into a big black hole takes a time much
shorter than ${\cal T}$.   

Towards its end ${\cal V}$ should approach  spherical symmetry: because there
is no rotation, and the elastic forces that could keep it aspherical will
succumb to gravitation as the black hole gnaws its way through it.   Thus the
mostly devoured system will not emit much in the way of gravitational
waves.  Finally, as the black hole finishes its meal, it recovers its
original Schwarzschild form, and establishes thermodynamic equilibrium with
the radiation filling the box.  In the presence of a
spherically symmetric system inside it, the box can be regarded as perfectly
spherical and thus gravitationally irrelevant (apart from the gravitational
redshift it induces inside it).

How much entropy in thermal gravitons $S_g$ leaks out of the box up to
the demise of ${\cal V}$ ?  As we saw, very little of
the black hole evaporates till equilibration with the box, so in this stage
gravitons carry away entropy $\ll S_{BH}=4\pi m^2/\ell_P{}^2$.  Now consider
the highly contrived situation that after equilibration,
accretion from ${\cal V}$ just balances the energy loss to Hawking gravitons.
The hole's mass $m$ then remains unchanged while it emits entropy at a rate
$1/(1920 m)$ [see Eq.~(\ref{power}) with $T_H=\ell_P{}^2/(8\pi m)$]. Were this
situation to endure over time ${\cal T} = 5120\pi m^3/(N\ell_P{}^2)$, the
entropy in gravitons would be $\sim m^2/(N\ell_P{}^2)$.  But since
${\cal V}$ transforms into a black hole in time much shorter than ${\cal T}$,
and the growth of the hole suppresses graviton emission, we may
safely conclude that $S_g=O(m^2/\ell_P{}^2)$. 

Let ${\cal E}$ be the equilibrium energy of all the radiation in the box. 
Then the final mass of the black hole is $<E+m-{\cal E}$ (spherical symmetry
allows us to just add masses).  We shall assume (and check presently) that
${\cal E}\ll E$;  hence the Hawking temperature is $T_H\approx
\ell_P{}^2/(8\pi E)$.  Again using Boltzmann's formulae to estimate ${\cal E}$
and the associated radiative entropy $S_b$ contained in the mostly flat
interior of the box with $r_0=2\cdot 10^2 E$, we have
\begin{equation}
{\cal E}={3T_H S_b\over 4}\approx {3\ell_P{}^2 S_b\over 32\pi
E}\approx{10^6 N\ell_P{}^2\over 5760\pi E}
\label{Boltzmann}
\end{equation}
The {\it relative\/} errors in these formulae are $O(m/E)+O({\cal
E}/E)$. Because $E>10^5 N^{1/2}\ell_P$, we get that for any $N$,  ${\cal E}/E<
10^{-8}$, as promised.

The GSL demands that the entropy of the final black hole plus $S_b$ plus
$S_g$ shall be no smaller than ${\cal S}$ plus the entropy of the initial
black hole.  Thus
\begin{equation}
{4\pi (E+m-{\cal E})^2\over \ell_P{}^2} + {32\pi E {\cal E}\over 3\ell_P{}^2}
+S_g > {\cal S} + {4\pi m^2\over \ell_P{}^2} 
\end{equation}
\label{balance}
Hence
\begin{equation}
{\cal S}<{4\pi E^2\over \ell_P{}^2}\times[1+2m/E+O(m^2/E^2)+O({\cal
E}/E)]
\label{bound}
\end{equation}
The corrections to unity are evidently small (recall that $m<10^{-1}E$).  We
have already remarked that ${\cal A}>4\pi(2E)^2$, where the inequality is {\it
not\/} marginal.  Hence it is clear that
\begin{equation}
{\cal S}< {{\cal A}\over 4\ell_P{}^2},
\end{equation}
which is the promised HB.  The universal entropy bound\cite{bek81} is very
similar to the HB for strongly self--gravitating systems, so to the extent
that circumscribing radius and proper energy are meaningful, it is also
established by this argument.

The derivation just given is appropriate only for strongly self--gravitating
systems.  A weakly self--gravitating system is much larger than its
gravitational radius, and this would necessitate making $r_0$ much larger than
we have assumed.  Because the energy in the box radiation scales as $r_0{}^3$,
were we to make $r_0$ larger, the energy in the cavity equilibrated with the
black hole would rapidly become comparable to $m$.  Of course we can still
treat systems on the borderline between strongly and weakly self--gravitating
by the above approach.  This allows us to bridge the gap between the present
approach and that given earlier for weakly self--gravitating systems.  
Systems {\it not\/} covered by our arguments are the strongly
self--gravitating solitons which have no particle like constituents, e.g.
\cite{bartnik}, and complexes of particles not obeying the constraint $\mu
E\gg 10^2\ell_P{}^2$.  The first exception is not very important: solitons
are vacuum solutions, and hence bear no entropy.

I thank Raphael Bousso and Mordehai Milgrom for
enlightening comments.

\end{document}